\documentclass[doublecol]{epl2} 

\usepackage{amsmath}

\title{The non-equilibrium part of the inertial range in
  decaying homogeneous turbulence}
\shorttitle{Non-equilibrium inertial range} 

\author{M. Obligado\inst{1} \and J.C. Vassilicos\inst{2}}
\shortauthor{M. Obligado \and J.C. Vassilicos}

\institute{                    
  \inst{1} Universit\'{e} Grenoble Alpes, CNRS, Grenoble-INP, LEGI, F-38000, Grenoble, France \\
  \inst{2} Univ. Lille, CNRS, ONERA, Arts et M\'etiers ParisTech, Centrale Lille, FRE 2017 - LMFL - Laboratoire de M\'ecanique des fluides de Lille - Kamp\'e de
Feriet, F-59000 Lille, France}


\abstract{We use two related non-stationarity functions as measures of
  the degree of scale-by-scale non-equilibrium in homogeneous
  isotropic turbulence. The values of these functions indicate
  significant non-equilibrium at the upper end of the inertial
  range. Wind tunnel data confirm Lundgren's (2002, 2003) prediction
  that the two-point separation $r$ where the second and third order
  structure functions are closest to their Kolmogorov scalings is
  proportional to the Taylor length scale $\lambda$, and that both
  structure functions increasingly distance themselves from their
  Kolmogorov equilibrium form as $r$ increases away from $\lambda$
  throughout the inertial range. With the upper end of the inertial
  range in non-equilibrium irrespective of Reynolds number, it is not
  possible to justify the Taylor-Kolmogorov turbulence dissipation
  scaling on the basis of Kolmogorov equilibrium.}

\begin{document}

\maketitle

\section{Introduction}

The Kolmogorov phenomenology relies on the presence of a statistically
stationary, i.e. equilibrium, cascade of energy where large-scale
energy input rate balances turbulence dissipation rate effectively
instantaneously. Assuming homogeneous or locally homogeneous
turbulence, and assuming external power input to be either absent or
limited to wavenumbers much smaller than $k$, the interscale energy
balance can be expressed as (see \cite{goto2016local,goto2016unsteady,
  sagaut2018homogeneous}),

\begin{equation}
{\partial \over \partial t} K^{>} (k,t) = \Pi (k,t) -\varepsilon^{>}(k,t)
\end{equation}

\noindent where $K^{>} (k,t)=\int_k^\infty E(k',t)dk'$ is the
high-pass filtered kinetic energy (with $E(k,t)$ the energy spectrum),
$ \Pi (k,t)$ is the interscale energy flux and $\varepsilon^{>}(k,t)$
is the high-pass filtered turbulence dissipation rate.
  
The Kolmogorov equilibrium hypothesis is $\vert {\partial \over
  \partial t} K^{>} (k,t)\vert \ll \varepsilon^{>}(k,t)$ (see
\cite{goto2016local,goto2016unsteady,tennekes1972first}). The size of
the largest turbulent eddies is typically accessed via the integral
length-scale $L$, and if this equilibrium hypothesis can be extended
to these largest eddies, i.e. to wavenumbers comparable to $2\pi/L$,
then

\begin{equation}
\varepsilon \approx \Pi (2\pi/L,t) \sim K^{3/2}/L
\end{equation}
  
\noindent where the scaling $K^{3/2}/L$ of $\Pi (2\pi/L,t)$ comes from
the expectation that there are no direct viscosity effects at the
largest scales and that the memory of initial/inlet conditions has had
time to fade away ($K$ is the turbulent kinetic energy). The resulting
scaling $\varepsilon = C_{\varepsilon} K^{3/2}/L$ where
$C_{\varepsilon} = Const$ (sometimes refered to as zeroth law of
turbulence or Taylor-Kolmogorov turbulence dissipation scaling) is
perhaps the most important consequence of the equilibrium cascade
phenomenology as it has a very wide range of implications in
turbulence theory and modeling (see \cite{tennekes1972first,
  vassilicos2015dissipation}).

However, a different dissipation scaling has been found over the past
ten years in various turbulent flows, including forced periodic
turbulence and decaying periodic turbulence
\cite{goto2016local,goto2016unsteady, goto2015energy}, various types
of grid-generated turbulence \cite{vassilicos2015dissipation} and
various turbulent shear flows \cite{vassilicos2015dissipation,
  obligado2016nonequilibrium, nedic2017dissipation, cafiero2019non}:

\begin{equation}
\varepsilon \sim U_{0} L_{0} K/L^{2},
\end{equation}

\noindent or equivalently $C_{\varepsilon}\sim Re_{I}/Re_{L} \sim
\sqrt{Re_{I}}/Re_{\lambda}$ where $R_{L} = \sqrt{K} L/\nu$,
$Re_{\lambda} =\sqrt{K}\lambda/\nu$ and $Re_{I} = U_{0} L_{0}/\nu$,
$U_0$ and $L_0$ being initial/inlet velocity and length scales
respectively ($\nu$ is the fluid's kinematic viscosity and $\lambda$
the Taylor length-scale). One can expect this scaling to be a
consequence of a cascade which is out of equilibrium, at least in the
upper part of the inertial range (closer to $L$), and this was indeed
confirmed by Goto \& Vassilicos (2016) \cite{goto2016unsteady} in
direct numerical simulations of freely decaying periodic
turbulence. These authors also found, in agreement with grid-generated
turbulence results (see \cite{vassilicos2015dissipation}), that the
classical dissipation scaling $\varepsilon \sim K^{3/2}/L$
($C_{\varepsilon} = Const$) appears rather suddenly after a number of
turnover times and as the local Reynolds numbers $R_{L}$ and
$Re_{\lambda}$ decrease. They also reported that this classical
dissipation scaling is not a reflection of a Kolmogorov equilibrium
cascade but in fact coexists with non-equilibrium over a wide range of
scales, including those inertial scales closer to $L$, as in
  the model of Bos et al \cite{bosetal2007}. Bos et al
  \cite{bosetal2007} explained this non-equilibrium version of
  $C_{\varepsilon} = Const$ by relating it to its equilibrium version
  and Goto \& Vassilicos (2016) \cite{goto2016unsteady} explained it
by noting that the three terms in equation (1) remain numerically
comparable as they decay in time, and proportional to each other for
$kL$ fixed close to $2\pi$.

The question arises whether this non-equilibrium, in particular at the
inertial scales closer to $L$, persists as the Reynolds number
increases towards infinity. This question is addressed in this paper
in the context of freely decaying homogeneous isotropic turbulence
(HIT) by using published wind tunnel data and Lundgren's
\cite{lundgren2002kolmogorov, lundgren2003kolmogorov} matched
asymptotic expansions for the second and third order structure
functions.

\section{Finite Reynolds number effects}

We work with the K\'arm\'an-Howarth equation for decaying HIT in a
rearranged version of the form found in \cite{landau1959fluid,
  danaila1999generalization}:


\begin{equation}
-\varepsilon F =  -{\left<(\delta u)^{3}\right>\over r} -{4\over
    5}\varepsilon +{6\nu\over r}{\partial \over \partial r} \left<(\delta
  u)^{2}\right>
  \end{equation}

\noindent where $F \equiv -{3\over \varepsilon r^{5}} \int_{0}^{r} dr'
r'^{4} {\partial \over \partial t} \left<(\delta u)^{2}\right>$ is the
non-stationarity/non-equilibrium term. $\left<(\delta u)^2\right>$ and $\left<(\delta
u)^3\right>$ are, respectively, the second and third order longitudinal
structure functions defined in terms of $\delta u =u(x+r,t)-u(x,t)$
where $u$ is the fluctuating velocity component in the direction of
the $x$ axis, and $r$ is the distance between points $x$ and $x+r$ on
that axis. This equation is the physical space equivalent of (1) where
$-\varepsilon F$ and $-{\left<(\delta u)^{3}\right>\over r}$ correspond to
${\partial \over \partial t} K^{>} (k,t)$ and $\Pi (k,t)$
respectively.

Finite Reynolds number (FRN) effects come from the
non-stationarity/non-equilibrium term $-\varepsilon F$ and from the
viscous term ${6\nu\over r}{\partial \over \partial r} \left<(\delta
u)^{2}\right>$ \cite{danaila1999generalization}. It has been proved
rigorously \cite{laizet2013interscale, valente2015energy} using simple
kinematic constraints on the basis that $K$ is finite that ${6\nu\over
  r}{\partial \over \partial r} \left<(\delta u)^{2}\right> \ll {4\over
  5}\varepsilon$ if $r\gg \lambda$. Therefore, the viscous term contributes no
significant FRN effects at scales $r\gg \lambda$.

We define the non-stationarity function $f$ by ${\partial \over
  \partial t} \left<(\delta u)^{2}\right> \equiv -\varepsilon f$. This
function provides a direct comparison between ${\partial \over
  \partial t} \left<(\delta u)^{2}\right>$ and $\varepsilon$ across
scales. It is related to the non-stationarity function $F$ by
$F={3\over r^{5}} \int_{0}^{r} dr' r'^{4} f$ and $f= {1\over 3 r^{4}}
{d\over dr} r^{5} F$. Note that both $f$ and $F$ are dimensionless and
non-negative.

It is easily seen that $f(r)\to 4/3$ and $F(r)\to 4/5$ in the limit
$r\gg L$ because $\left<(\delta u)^{2}\right> \to {4\over 3}K$ in that limit.

From a first order Taylor expansion and $\varepsilon = 15\nu
\left<({\partial u\over \partial x})^{2}\right>$, $\left<(\delta
u)^{2}\right> \approx r^{2}{\varepsilon\over 15\nu}$ for $r\ll
\lambda$ \cite{taylor1935statistical}. Differentiating both sides with
respect to time, we obtain $f \approx \left({r\over
  \lambda}\right)^{2}{\lambda^{2}\over 15\nu}\left(-{{d\over
    dt}\varepsilon \over \varepsilon}\right)$ which implies $f \approx
Const ({r\over \lambda})^{2}$ for $r\ll \lambda$ if $K$ decays as a
power of time; $F \approx {3\over 7} Const ({r\over \lambda})^{2}$
follows under these same conditions.

We can expect the non-stationarity functions $f$ and $F$ to increase
from $0$ at $r=0$ to $4/3$ and $4/5$, respectively, at $r\gg
L$. Incidentally, if $f$ is a monotonically increasing function of
$r$, so is $F$. (Proof: ${\partial \over \partial r} F >0$ if $r^{5}f
> 5 \int_{0}^{r} dr' r'^{4} f(r')$ which follows from $5 \int_{0}^{r}
dr' r'^{4} f(r') < 5 \int_{0}^{r} dr' r'^{4} f(r) = r^{5}f$ given that
$f(r)> f(r')$ for $r>r'$.)

We might expect a tendency towards equilibrium at the smaller inertial
scales, but what are the values of $f$ and $F$ for $r$ smaller than
but comparable to $L$? The answer to this question can inform us about
the degree of non-equilibrium at the larger inertial scales.

\section{Experimental data} We use published wind tunnel
measurements of $\left<(\delta u)^{2}\right> (r)$ and $\left<(\delta
u)^{3}\right> (r)$ for approximate HIT to calculate $F(r)$ from (4)
and then $f(r)$ from $f= {1\over 3 r^{4}} {d\over dr} \left(r^{5} F
\right)$. The only published data sets that we could find for wind
tunnel measurements of both $\left<(\delta u)^{2}\right>$ and
$\left<(\delta u)^{3}\right>$ in approximate HIT conditions are from
the grid turbulence experiments of Zhou \& Antonia
\cite{zhou2000reynolds}, Malecot \cite{malecot1998intermittence} and
Bourgoin et al \cite{bourgoin2018investigation} and from the Modane
wind tunnel measurements of \cite{gagne2004reynolds}. Note that the
data of Bourgoin et al \cite{bourgoin2018investigation} have been
obtained in the same Modane wind tunnel as \cite{gagne2004reynolds}.
The particular datasets that we selected from these publications span
the widest possible range of Reynolds numbers based on
$Re_\lambda$. Table \ref{tab1} summarises the turbulence parameters
for each dataset.


All these data have been obtained with hot-wire anemometry (the
specific details of the experimental set-ups for each value of
$Re_\lambda$ are given in the corresponding references), although for
the two larger values of $Re_\lambda$ the Kolmogorov length scale
$(\eta=(\nu^3/\varepsilon)^{1/4})$ was not fully resolved. In all
cases, the conversion from time to space was done via the Taylor
hypothesis. The Taylor length-scale is defined from
$\lambda^2=<u^2>/<(\partial u / \partial x)^2>$, and $Re_\lambda$ is
calculated as $Re_\lambda= u'\lambda /\nu$ where $u' \equiv
<u^2>^{1/2}$. The turbulence dissipation rate, assuming HIT, can be
obtained from $\varepsilon=15 \nu <u^2>/\lambda^2$. For the two larger
values of $Re_\lambda$, where $\eta$ was not resolved, $\varepsilon$
and $\lambda$ have been obtained via the compensated second order
structure function $\frac{{\left<(\delta u)^2 \right>}}{(\varepsilon
  r)^{2/3}}$ ($Re_{\lambda}=380$) or by matching the shape of $\frac{{\left<(\delta u)^3
\right>}}{(\varepsilon r)}$ at $r$ close to $\eta$ to other data
($Re_{\lambda}=2260$). Finally, the lengthscale
$L$ is calculated from the autocorrelation $R_{uu}$ as $L=\int_0^{\infty}
R_{uu} dr$.

\begin{table*}[h!]
\begin{center}
\begin{tabular}{|c|c|c|c|c|c|c|c|}
Grid & $Re_\lambda$ & $U_\infty$ (m/s) & $L$ (cm) & $\eta$ (mm) &
$\varepsilon$ (m$^2$s$^{-3}$) & $\lambda$ (cm) & $u'$ (m/s)\\ \hline
\hline ZA1 & 50 & 4.8 & 5.52 & 0.58 & 0.032 & 0.81 & 0.096\\ ZA2 & 75
& 11.3 & 4.21 & 0.30 & 0.40 & 0.54 & 0.22\\ ZA3 & 89 & 6.7 & 3.81 &
0.24 & 1.37 & 0.41 & 0.325\\ Ma & 144 & 16.9 & 4 & 0.15 & 7.7 & 0.35 &
0.64 \\ MI & 380 & 34.2 & 22.0 & 0.23 & 1.19 & 0.86 & 0.63\\ Mo & 2260
& 20.75 & $\approx$ 410 & 0.30 & 0.95 & 2.80 & 1.58\\
\end{tabular}
\end{center}
\caption{Experimental parameters : Reynolds number based on Taylor
  micro-scale ($Re_\lambda$), incoming wind velocity ($U_\infty$),
  integral length scale ($L$), Kolmogorov scale ($\eta$), energy
  dissipation rate ($\varepsilon$), Taylor micro-scale ($\lambda$) and
  standard deviation of fluctuating streamwise velocity ($u'$). $ZA$
  stands for Zhou \& Antonia \cite{zhou2000reynolds}, $Ma$ for
  Malecot \cite{malecot1998intermittence}, $MI$ for the Modane
  inflatable grid-turbulence experiment
  \cite{bourgoin2018investigation} and $Mo$ for the Modane
  measurements of Gagne et al \cite{gagne2004reynolds}}\label{tab1}
\end{table*}

\section{Results} 
\begin{figure}
\setlength{\unitlength}{1cm}
\begin{center}
\includegraphics[width=0.5\textwidth,keepaspectratio]{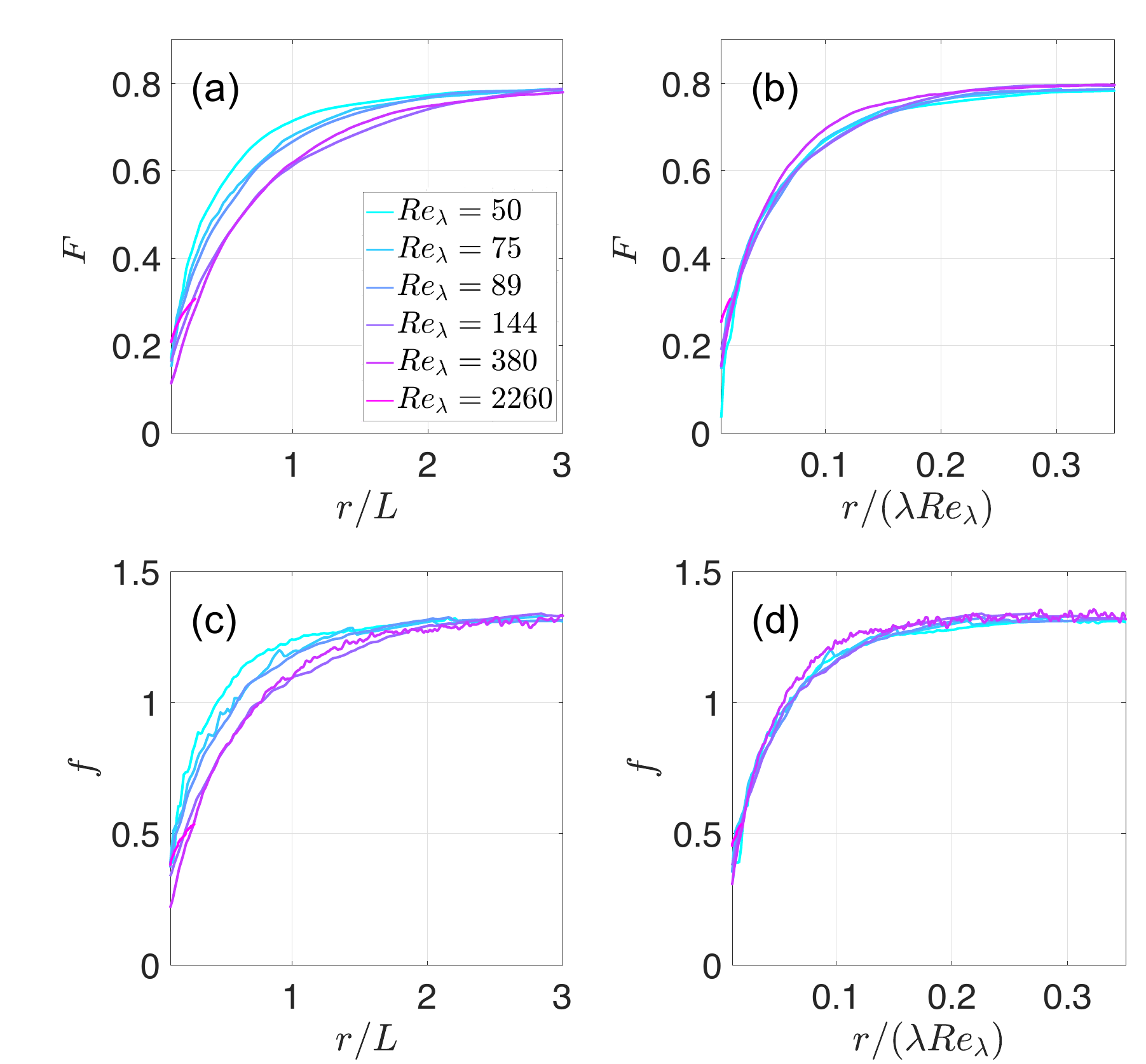}\\
\end{center}
\caption{Top plots: non-stationarity function $F$ vs (a) $r/L$ and (b)
  $r/(\lambda Re_\lambda)$. Bottom plots: non-stationarity function
  $f$ vs (c) $r/L$ and (d) $r/(\lambda Re_\lambda)$.  The colour code
  for $Re_\lambda$ indicated in (a) is the same in all the plots in
  this paper and the data sets used to obtain $F$ and $f$ are described
  in Table 1.}
\label{fig1}
\end{figure}

In figure \ref{fig1} we plot the non-stationarity
functions $F$ and $f$ as functions of $r/L$ and $r/(\lambda
Re_\lambda)$. Note the relation $15 L = C_{\varepsilon} \lambda
Re_{\lambda}$ in HIT, hence $\lambda Re_{\lambda}$ represents the
integral scale too but without FRN effects coming from
$C_{\varepsilon}$. $F$ is found to be larger than 0.2 and $f$ larger
than 0.4 for all $r$ larger or equal to $L/5$ in all data sets. In the
case of our highest Reynolds number, $Re_{\lambda}=2260$,
$r=40\lambda$ corresponds to $r\approx L/4$, which shows that $L/5$ is
quite a small scale for all our data and well within the smallest
scales of the inertial range for most if not all of them. We must
conclude that most if not all the inertial range is not in Kolmogorov
equilibrium in all our data sets, and definitely not in the part of
the inertial range closer to its upper bound $L$.  Justifying
$C_{\varepsilon} = Const$ on the basis of such equilibrium at the
Reynolds numbers considered here is definitely questionable.

Note the much better collapse with $\lambda Re_{\lambda}$ than with
$L$ for both $F$ and $f$ in figure \ref{fig1}. This might suggest that
the lack of collapse with $L$ is caused by the residual Reynolds
number dependence of $C_{\varepsilon}$ at the smaller $Re_{\lambda}$
values. Extrapolating to $Re_{\lambda}$ larger than $2260$ might
suggest that $F$ and $f$ remain larger than 0.2 and 0.4, respectively,
for $r\ge L/5$ as $Re_{\lambda}$ grows beyond $2260$. A
non-stationarity function $f$ larger than 0.4 means that $-{\partial
  \over \partial t} \left<(\delta u)^{2}\right> > 0.4
\varepsilon$. Hence, the current state of experimental evidence for
HIT does not rule out, and in fact might even suggest, that the upper
part of the inertial range between $L$ and an order of magnitude
smaller than $L$ is out of equilibrium at all Reynolds numbers. This
conclusion is supported by Lundgren's \cite{lundgren2002kolmogorov,
  lundgren2003kolmogorov} matched asymptotic expansions for the second
and third order structure functions which we examine next.

\subsection{Extrapolation to even higher $Re_{\lambda}$}
Lundgren \cite{lundgren2002kolmogorov, lundgren2003kolmogorov}
obtained the following forms for $\left<(\delta u)^{2}\right>$ and
$\left<(\delta u)^{3}\right>$ from the Navier-Stokes equation, and in
particular the K\'arm\'an-Howarth equation, by using the method of
matched asymptotic expansions in the range $\eta \ll r\ll L$:
  
  \begin{equation}
  {\left<(\delta u)^{2}\right>\over (\varepsilon r)^{2/3}} \approx C [1 - A_{2}
    (r/L)^{2/3} -B_{2} (r/\eta)^{-4/3}]
\end{equation}
  \begin{equation}
  -{\left<(\delta u)^{3}\right>\over (\varepsilon r)} \approx 4/5 - A_{3}
  (r/L)^{2/3} -B_{3} (r/\eta)^{-4/3}
  \end{equation}

\noindent where $A_{2}$, $B_{2}$, $A_{3}$ and $B_{3}$ are
dimensionless constants of order $1$, and $C=2$ from widely accepted
experimental evidence to date (e.g. see Pope 2000 \cite{pope2000turbulent}).
It is worth mentioning that these corrections to the
  Kolmogorov equilibrium scalings of $\left<(\delta u)^{2}\right>$ are
  different from the non-equilibrium correction obtained by Yoshizawa
  \cite{yoshizawa1994} from his spectral closure theory and by Bos \&
  Rubinstein \cite{BosRubinstein2017} who obtained the same correction
  as Yoshizawa by using Kovaznay’s spectral closure model. Whilst the
  correction of these authors scales as $r^{2/3}$ similarly to the
  correction $A_{2}(r/L)^{2/3}$ in (5), it also scales with
  $d\varepsilon/dt$ which is not generally the case in (5) (unless we
  assume, for example, that the $K-\varepsilon$ equation for
  $d\varepsilon/dt$ and $C_{\varepsilon} = Const$ hold true). More
  importantly, however, the non-equilibrium correction of Yoshizawa
  and Bos \& Rubinstein does not capture the $B_{2} (r/\eta)^{-4/3}$
  correction in (5) which is essential for our conclusions as the
  special role of the Taylor microscale described below cannot be
  captured without it.

Using (1), formulae (4) and (5) lead to

\begin{equation}
\begin{aligned}
F \approx & [A_{3} (r/L)^{2/3}- 4C Re_{\lambda}^{-2}(r/L)^{-4/3} + \\
&    8CA_{2} Re_{\lambda}^{-2} (r/L)^{-2/3}] - 
    \\ & B_{3}   Re_{\lambda}^{-2/3}(r/\lambda)^{-4/3}[1 + 4C{B_{2}\over B_{3}}
    Re_{\lambda}^{-2}(r/L)^{-4/3}].
    \end{aligned}
\end{equation}
 
\noindent This expression for $F$ implies that $F \approx A_{3}
(r/L)^{2/3}$ in the limit $Re_{\lambda}\to \infty$ as $r/L$ is kept
constant or as $r$ and $t$ are both kept constant relative to
laboratory length and time scales. In other words, the higher part of
the inertial range does not tend towards equilibrium as
$Re_{\lambda}\to \infty$, it keeps its level of non-equilibrium
unchanged.

\begin{figure}
\setlength{\unitlength}{1cm}
\begin{center}
\includegraphics[width=0.5\textwidth,keepaspectratio]{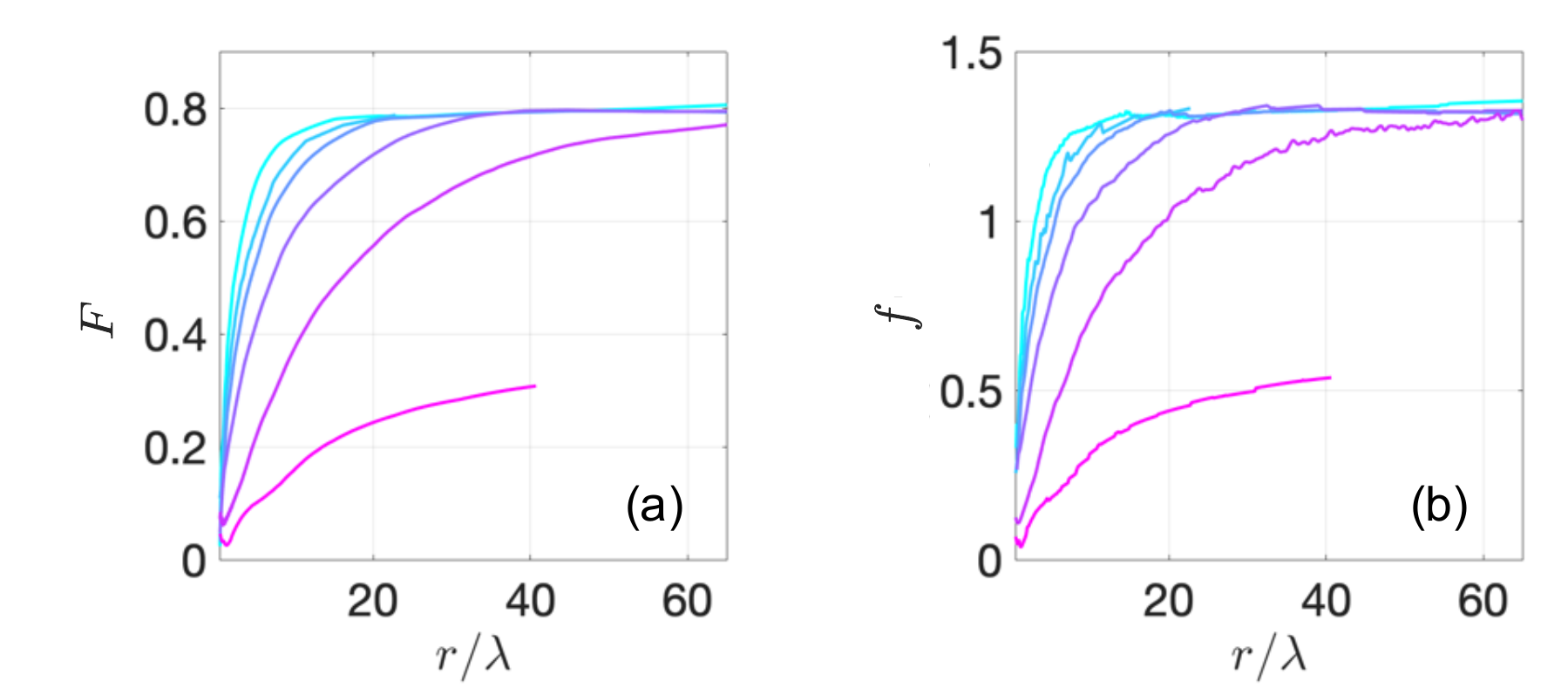}\\
\end{center}
\caption{Functions (a) $F$ and (b) $f$, both plotted vs
  $r/\lambda$. The colour code for different $Re_\lambda$ is the same
  as in figure 1 and the data sets used to obtain $F$ and $f$ are
  described in Table 1.}
\label{fig2}
\end{figure} 

This observation does not contradict the fact that, as
$Re_{\lambda}\to \infty$ and $L/\lambda \to \infty$, an approximate
equilibrium range does develop in the limit where $r/L \to 0$. Indeed,
for fixed $r/\lambda$, $F$ does tend to $0$ as $Re_{\lambda}\to
\infty$, and our data show evidence in agreement with this in figure
\ref{fig2}: note how both $F$ and $f$ decrease, presumably
  towards zero, as $Re_\lambda$ increases while keeping $r/\lambda$
  constant. However, this asymptotic equilibrium part of the inertial
range is so far from $L$, and increasingly so with increasing
$Re_{\lambda}$, that it cannot be used as a basis for $C_{\varepsilon}
= Const$.

\begin{figure}
\setlength{\unitlength}{1cm}
\begin{center}
\includegraphics[width=0.5\textwidth,keepaspectratio]{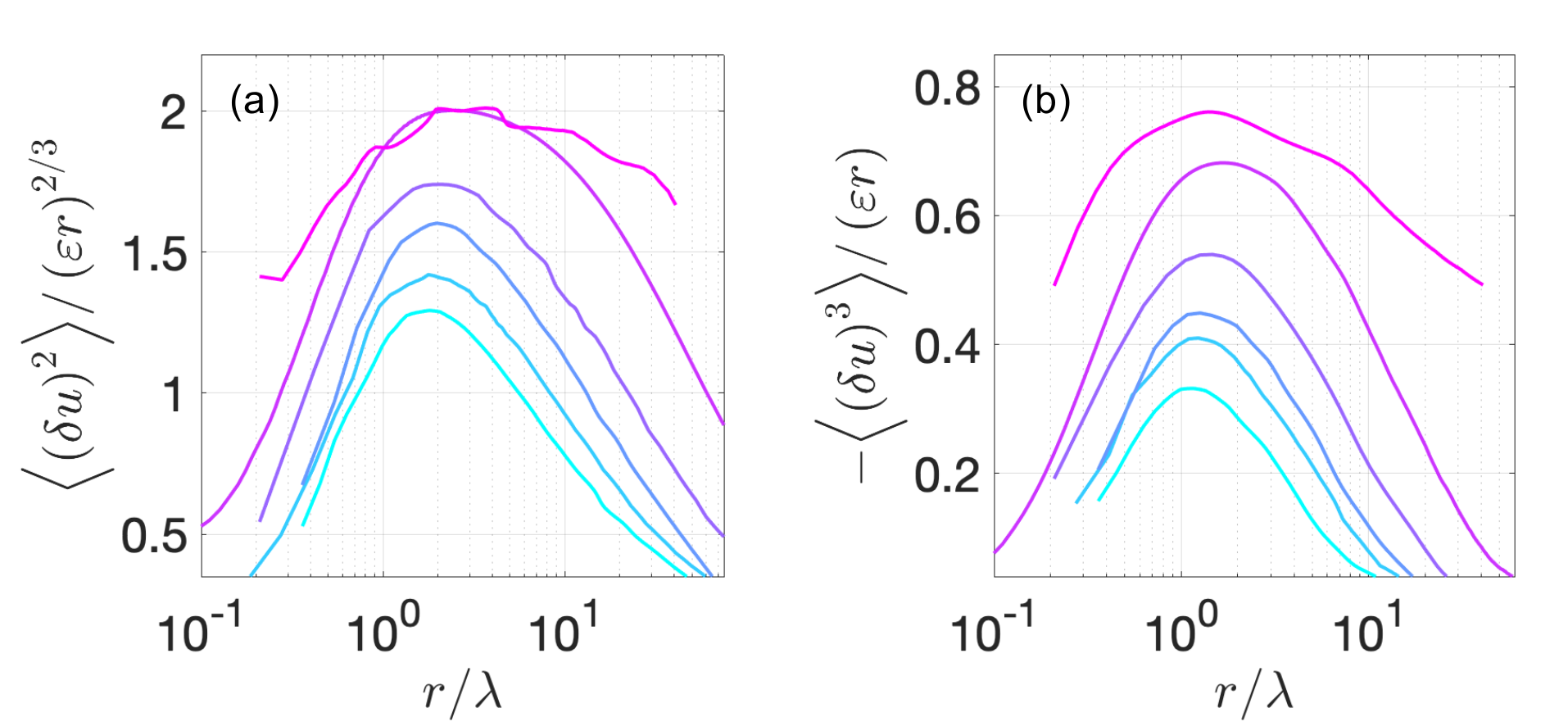}\\
\end{center}
\caption{Compensated (a) second-order structure function
  $\frac{\left<{(\delta u)^2}\right>}{(\varepsilon r)^{2/3}}$ and (b)
  third-order structure function $\frac{\left<{(\delta
      u)^3}\right>}{\varepsilon r}$, both plotted vs $r/\lambda$.}
\label{fig3}
\end{figure}

To clarify this distinction between the asymptotic equilibrium scales
at the lower end of the inertial range and the non-equilibrium scales
at the upper end of the inertial range we calculate the value
$r_{max}$ of $r$ where Lundgren's formulae for ${<(\delta u)^{2}>\over
  (\varepsilon r)^{2/3}}$ and $-{\left<(\delta u)^{3}\right>\over
  (\varepsilon r)}$ have a maximum. As already noted in
\cite{lundgren2003kolmogorov}, the maximum is at $r_{max} \sim
\lambda$ for both ${\left<(\delta u)^{2}\right>\over (\varepsilon
  r)^{2/3}}$ and $-{\left<(\delta u)^{3}\right >\over (\varepsilon
  r)}$. Our data confirm this prediction and give $r_{max} \approx
2\lambda$ for ${\left<(\delta u)^{2}\right>\over (\varepsilon
  r)^{2/3}}$ and $r_{max} \approx 1.5\lambda$ for $-{\left<(\delta
  u)^{3}\right>\over (\varepsilon r)}$, as shown in figures \ref{fig3}a\&b. To our knowledge, this is the first time that this
prediction is confirmed for the second order structure function and
the first time that it is confirmed over such a wide range of Reynolds
numbers for $-{\left<(\delta u)^{3}\right>\over (\varepsilon r)}$
(data from only two Reynolds numbers from Yves Gagne's Modane
measurements were used in \cite{lundgren2003kolmogorov}).

It may be worth pointing out that the empirical formula for
$\left<(\delta u)^{2}\right>$ in \cite{kurien2000anisotropic,
  antonia2006approach}, namely
 \begin{equation}
  {\left<(\delta u)^{2}\right>\over \left(\varepsilon \eta \right)^{2/3}} = {(r/\eta)^{2}
    (1+r/L)^{-2/3}\over 15 [1 + D(r/\eta)^{2}]^{2/3}}
 \end{equation}
 where $30 D^{2/3}=1$, leads to a maximum of ${\left<(\delta
   u)^{2}\right>\over \left(\varepsilon r \right)^{2/3}}$ at $r_{max}
 \approx 2^{5/6} C_{\varepsilon} \lambda$ for $L/\eta \gg 1$ (using
 $\varepsilon = C_{\varepsilon} u'^{3}/L = 15 \nu u'^{2}/\lambda^{2}$
 to relate $\lambda$ to $L$ via $15L/\lambda = C_{\varepsilon}
 Re_\lambda$). This empirical formula is designed by construction to
 give $<(\delta u)^{2}> \approx 2(\varepsilon r)^{2/3}$ in the range
 $\eta \ll r\ll L$, and the right kinematic behaviours at $r/L\gg 1$
 and $r/\eta \ll 1$. It is interesting that it combines dependencies
 on $L$ and $\eta$ in exactly the right way for ${\partial \over
   \partial r} {\left<(\delta u)^{2}\right>\over \left(\varepsilon r
   \right)^{2/3}} =0$ to yield $r_{max} \sim \lambda$, in agreement
 with Lundgren's prediction.

\section{Conclusions} Given that the inertial range is, by definition,
a range where the viscosity is effectively absent, it must be defined
as $\lambda \ll r \ll L$. Our data and Lundgren's
\cite{lundgren2002kolmogorov, lundgren2003kolmogorov} formulae (5) and
(6) show that $\left<(\delta u)^{2}\right>$ and $\left<(\delta
u)^{3}\right>$ are closest to their Kolmogorov equilibrium predictions
at the lower end of the inertial range, i.e. $r=r_{max} \sim
\lambda$. Both $\left<(\delta u)^{2}\right>$ and $\left<(\delta
u)^{3}\right>$ increasingly deviate away from their Kolmogorov
equilibrium prediction as $r$ increases in the inertial range. The
matched asymptotic expansions (5) and (6) imply that $\left<(\delta
u)^{2}\right>$ and $\left<(\delta u)^{3}\right>$ tend towards
$C(\varepsilon r)^{2/3}$ and $-{4\over 5}\varepsilon r$, respectively,
as $Re_{\lambda}\to \infty$ for fixed $r/\lambda$ but not for fixed
$r/L$.  For fixed $r/L$ smaller that 1, they tend to

  \begin{equation}
\left<(\delta u)^{2}\right> = C (\varepsilon r)^{2/3} [1 - A_{2}
  (r/L)^{2/3}]
  \end{equation}

  \noindent
  and
  \begin{equation}
  \left<(\delta u)^{3}\right> = -(\varepsilon r)[ 4/5 - A_{3}
  (r/L)^{2/3}].
  \end{equation}

\noindent
The upper end of the range is therefore always significantly out of
equilibrium whatever the Reynolds number. The wind tunnel data that we
have compiled here for HIT support this conclusion and suggest that
the upper end of the inertial range where $r$ is up to one order of
magnitude smaller than $L$ is such that $-{\partial \over \partial t}
\left<(\delta u)^{2}\right>$ is comparable to $\varepsilon$. Any
justification of $C_{\varepsilon} = Const$ in terms of Kolmogorov
equilibrium is therefore, at the very least, seriously questionable in
decaying HIT, at any Reynolds number.


\acknowledgments

This work was motivated by a discussion with Claude Cambon during a
dinner held in his honour in Lyon to mark his 65th year.

\end{document}